# A Robust Data-Driven Fault Diagnosis scheme based on Recursive Dempster–Shafer Combination Rule*

N. Cartocci, M.R. Napolitano, G. Costante, F. Crocetti, P. Valigi, M.L. Fravolini

*Abstract*— In-flight sensor fault diagnosis and recursive combination of residual signals via the Dempster-Shafer (DS) theory have been considered in this study. In particular, a novel evidence-based combination rule of residual errors as a function of a reliability measure derived from streaming data is proposed for the purpose of online robust sensors fault diagnosis. The proposed information fusion mechanism is divided into three steps. In the first step, the classic DS probability mass combination rule is applied; then, the difference between the previous posterior mass and the current prior mass associated with fault events is computed. Finally, the increment of the posterior mass of a fault event is weighted as a function of a reliability coefficient that depends on the norm of control activity. A Sensor Fault Isolation scheme based on the proposed combination rule has been worked out and compared with well-known state-of-the-art recursive combination rules. A quantitative analysis has been performed exploiting multi-flight data of a P92 Tecnam aircraft. The proposed approach showed to be effective, particularly in reducing the false alarms rate.

## I. INTRODUCTION

Fault Diagnosis plays an important role in accident prevention, human safety, maintenance, decision-making, and cost minimization [1]. Fault diagnosis has been extensively applied to improve the safety and reliability for a wide range of engineering systems such as mechanical, chemical, nuclear, and electric networks [2]. Fault Diagnosis methodologies can be conceptually divided into two main categories: model-based and data-based. The model-based approaches rely on mathematical models derived from the physical laws governing the dynamics of the system and are used to derive diagnostic signals that are sensitive to specific faults. Data-based approaches, instead, derive diagnostic signals from experimental models identified directly from data acquired from the plant. The latter approaches are preferable in case a detailed physical knowledge of the system is not available or when system input-output relations are too complex or uncertain [3]. Nowadays multivariate Statistical Process Monitoring (SPM) methods, in particular, Principal Component Analysis (PCA) [3], and directional residuals [4], [5] are popular tools used for process monitoring. The success of PCA based techniques lies in the ease of use and in the capacity to efficiently manage many variables and big data. However, the performance of these techniques depends strongly on the number of the selected principal components, on the statistical indices used (mainly SPE and T2) and on the corresponding Fault Detection (FD) and Fault Isolation (FI) thresholds. Considering noisy and correlated experimental residuals, it is often difficult, and time-consuming, the set-up of a reliable fault diagnosis logic because the above indices are extremely sensitive to the noise and modelling uncertainty that may produce a significant number of false positives and false negatives in practical applications. A basic approach for the FI is the so-called contribution plots method (also known as Reconstruction Based Contributions method) [3]. Although simple, this method may produce wrong fault isolations due to the smearing effect, that is the influence of a faulty sensor measurement to the contributions associated to non-faulty sensors [6]. In the case of a limited number of monitored variables, directional residuals methods have proven to be a valid and reliable alternative to SPM based methods [7].

Considering safety-critical systems, such as aircraft, a reliable in-flight condition monitoring is essential to limit false alarms. Robust performance can be achieved by considering in the diagnostic algorithm not only the information of the current state of the aircraft but also the temporal history of the sensor measurements [3]–[5]. To improve the robustness of fault diagnosis techniques these are often used in conjunction with Evidence-Based Filtering (EBF) that propagate fault probability information through time with the purpose of improving alarms consistency. Traditional information fusion methods are Drift-Diffusion Models (DDMs) [3], Bayesian filtering [4], [5], Dempster-Shafer Theory (DST) and their evolutions. DDMs assume that decisions are taken as a result of a sequential process that accumulates sensorial information over time until boundary thresholds are reached. Bayesian filtering is used to estimate the probability of an event basing on prior information applying the Bayes' theorem. The latter approach has some disadvantages compared to Dempster-Shafer inference methods. Among these problems, Bayesian filtering requires the knowledge of prior probability distributions that may not be available [8]; in contrast, DST allows information fusion via a simple reasoning evidence-based process that does not require the knowledge of prior probabilities distributions [9]. Although DST has the mentioned advantage, this may generate paradoxical results when used to fuse highly conflicting evidence information [10], further, in case it is used recursively, the accumulated evidence may suddenly saturate toward the boundary values (that is accumulated evidence saturates toward 1 or 0) [11], [12].

To date, few studies have been presented in the literature to face the problems originating from the sequential application

*Research supported by projects: Clean Sky 2 Joint Undertaking (CS2JU) under the European Union's Horizon 2020 research and innovation program, under project no. 821079 E-Brake; and by the project Basic Research Funds 2018 University of Perugia (Project: RICBA18MR)

N. Cartocci, G. Costante, F. Crocetti, P. Valigi, M.L. Fravolini are with the Department of Engineering, University of Perugia, Perugia, 06125 Italy (e-mails: nicholas.cartocci@unipg.it, gabriele.costante@unipg.it, paolo.valigi@unipg.it, francesco.crocetti@unipg.it, mario.fravolini@unipg.it).

M.R. Napolitano is with the Department of Mech. and Aerospace Engineering, West Virginia University, Morgantown, WV, 26506, USA (email: marcello.napolitano@mail.wvu.edu).

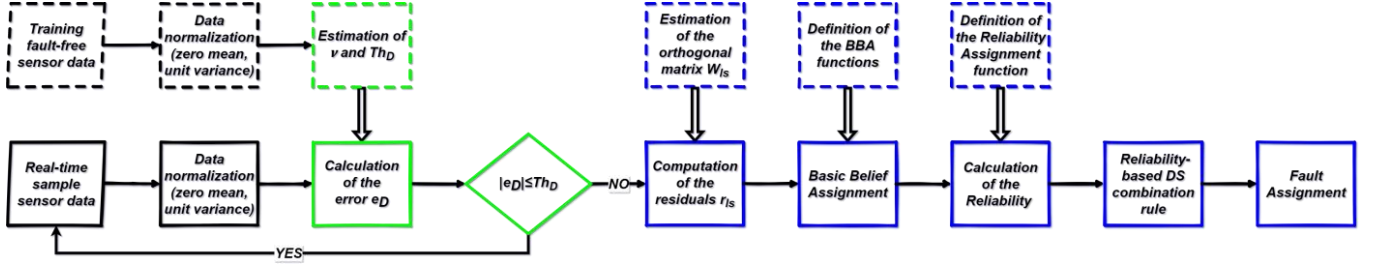

Figure 1. *The workflow of the proposed technique*

of the DS inference mechanism. Between these, two major methodologies have been proposed to manage sequential DS inference. One method preprocesses and redistributes the Basic Belief Mass (BMM), and the other modifies the basic DS combined rule [10] as reported hereafter. Powell and Roberts in [13] proposed a recursive algorithm based on the arithmetical mean of the conjunctive and disjunctive rule of combination where the BBM is discounted depending on the degree of precision (educatedness). Smarandache and Dezert in [14] developed six versions of the Proportional Conflict Redistribution where the DS inherent contradiction is avoided. Finally, Khan and Anwar in [12] have modified the original BBM of classic DST through an entropy measure.

In the present paper, we present a sensor fault diagnosis scheme based on robust directional residuals proposing a modification of the classical DS combination rule that takes into account in the fusion of the diagnostic residual information of the reliability level of the incoming measurements. The proposed combination rule, as long as well-known state of the art sequential DS information fusion methods, have been applied to manage the information produced by directional residuals signals produced by a sensor Fault Diagnosis system of an aircraft. The comparative analysis has been performed using multi-flight experimental data of a P92 Tecnam aircraft [15].

## II. OVERVIEW OF THE PROPOSED FDI APPROACH

The block diagram of the proposed FDI scheme is shown in Figure 1; this can be conceptually divided into two parts: Fault Detection (Green) and Fault Isolation (Blue). The Fault Detection (FD) part is based on the Singular Value Decomposition (SVD) and the concept of Right Null Space (RNS) of the experimental (Training) data. The Fault Isolation (FI) part is the heart of the proposed technique and is based on a novel Reliability-based Dempster-Shafer combination rule. The dashed blocks are associated with the off-line design that, in turn, is based on the experimental multi-flight data; the solid blocks, instead, are associated with the FDI operations that are performed in-flight. Models, parameters, and thresholds computed in the design phase are used in the online operation mode.

## III. SENSOR FAULT DIAGNOSIS MODEL

The set of the monitored (potentially faulty) sensors measurements is concatenated in the vector $x(k) \in \mathbb{R}^{n_x}$, while the set of control signals, as long other no monitored sensors (assumed not-faulty), is identified by the vector $u(k) \in \mathbb{R}^{n_u}$. The overall set of sensors is identified through the vector $z(k) = [x(k); u(k)] \in \mathbb{R}^n$, where $n = n_x + n_u$, and the integer $k$ is the sequence index at sample time $t = \Delta t\, k$ (where $\Delta t$ is the sampling interval). The proposed technique is based on Analytical Redundancy (AR) concepts and estimates the sensor measurements in $x(k)$ as a function of the vector $z(k)$. It is assumed that $x(k)$ is expressed as the sum of two contributions that is a linear multivariate model plus an uncertain term that characterizes modeling uncertainties, nonlinearities, and noise [4] that is:

$$x_i(k) = \sum_{\substack{j=1 \\ j \neq i}}^{n_x} w_{xi,j} x_j(k) + \sum_{j=1}^{n_u} w_{ui,j} u_j(k) + \Delta_i(k), \quad (1)$$
$$i = 1, \dots, n_x$$

where $w_{xi,j}$ and $w_{ui,j}$ are the coefficients of the linear multivariate model, and $\Delta_i(k)$ characterizes nonlinearities, uncertainty and noise associated with the $i$-th sensor. For simplicity, the models in (1) are rearranged as:

$$x_i(k) = w_{xi}\, x(k) + w_{ui}\, u(k) + \Delta_i(k), \quad i = 1, \dots, n_x \quad (2)$$

where $w_{xi} = [w_{xi,1}, \dots, w_{xi,i-1}, 0, w_{xi,i+1}, \dots, w_{xi,n_x}] \in \mathbb{R}^{n_x}$ and $w_{ui} = [w_{ui,1}, \dots, w_{ui,n_u}] \in \mathbb{R}^{n_u}$. Putting the above $n_x$ equations together, we get the following vector expression:

$$x(k) = W_x x(k) + W_u u(k) + \Delta(k) \quad (3)$$

where $W_x = [w_{xi}; \dots; w_{xn_x}] \in \mathbb{R}^{n_x \times n_x}$ and $W_u = [w_{ui}; \dots; w_{un_u}] \in \mathbb{R}^{n_u \times n_u}$ are constant matrices. The linear terms in (3) provide a linear (computable) estimation of $x(k)$ that is defined as:

$$\hat{x}(k) = W_x x(k) + W_u u(k) \quad (4)$$

Equation (4) can be expressed in a more compact form defining the matrices $W' = W_x - I \in \mathbb{R}^{n_x \times n_x}$ and $W = [W'\ W_u] \in \mathbb{R}^{n_x \times n}$:

$$W' x(k) + W_u u(k) = 0 \quad (5)$$
$$W z(k) = 0 \quad (6)$$

### A. Sensor fault modeling

In this study, single additive sensor faults have been considered. Fault on the $i$-th sensor is named as $f_i(k)$ and is summed to the fault-free signal $x_i(k)$. In the presence of a sensor fault, the vector $x(k)$ is replaced by its faulty version that is:

$$x(k) \leftarrow x(k) + F(k) \quad (7)$$

where $F(k) = [0 \dots 0\ f_i(k)\ 0 \dots 0]^T \in \mathbb{R}^{n_x}$ is the fault vector in case of a fault on the $i$-th sensor. Equation (5) with the injection of the fault becomes

$$r(k) = W'[x(k) + F(k)] + W_u u(k) \quad (8)$$

where $r(k)$ is commonly known as the primary residuals vector. Considering (5), the previous equation (8) is rearranged as:

$$r(k) = W' F(k) = w_i f_i(k) + \Delta(k) \quad (9)$$

where $w_i$ corresponds to the $i$-th column vector of $W'$ (and of $W$). Vector $w_i$ is known as the fault signature or the fault direction associated with fault $f_i(k)$. Assuming that $\Delta(k)$ is small (compared to fault amplitude), the vector $r(k)$ can be assumed parallel to the vector $w_i$. This directional information will be exploited, later, for sensor FI purposes.

## IV. Fault Detection

In this study, a robust Fault Detection (FD) method is developed that is based on the approach proposed in [16], [17] using an anomaly-sensitive versor $v \in \mathbb{R}^n$ such that:

$$e_D(k) = z(k)^T v = 0 \quad (10)$$

in fault-free conditions, and

$$e_D(k) = z(k)^T v = v_i f_i(k) \neq 0 \quad (11)$$

in case of a fault on the $i$-th sensor. Equation (11) is known in the literature as a parity equation [18], [19], while the error signal $e_D(k)$ is defined as the *detection residual* or the *detection error*. If the error $e_D(k)$ exceeds a predefined detection threshold ($Th_D$), then a fault condition on one (or more) sensors is detected:

$$\begin{aligned} |e_D(k)| \leq Th_D &\Rightarrow \textbf{Normal} \\ |e_D(k)| > Th_D &\Rightarrow \textbf{Fault Detection} \end{aligned} \quad (12)$$

The versor $v$ is determined performing Singular Value Decomposition of the (Training) data matrix $Z$ [20], [21], that is:

$$Z = U \Lambda V^T \quad (13)$$

where $U \in \mathbb{R}^{m \times m}$ is the left singular matrix, $V \in \mathbb{R}^{n \times n}$ is the right singular matrix, $\Lambda \in \mathbb{R}^{m \times n}$ is a rectangular diagonal matrix where the diagonal entries are known as the singular values of $Z$. As described in [22] there exists a versor $v \in \mathbb{R}^n$ belonging to the Right Null Space (RNS) of $Z$ and defined as the last column vector of $V$ is such that $Zv \approx 0$. In other words, $v$ is the eigenvector associated to the smallest eigenvalue of the data covariance matrix (the eigenvector associated to the direction of the smaller explained variance). The detection threshold $Th_D$ was derived from the experimental Cumulative Distribution Function (CDF) of $|e_D(k)|$ applying the procedure proposed in [23], where:

$$Th_D = min(|e_D(k)|): \ CDF(|e_D(k)|) \geq (1 - P_F) \quad (14)$$

where the $P_F$ is the probability that $|e_D(k)|$ exceeds the detection thresholds in fault-free condition and $CDF(|e_D(k)|)$ is defined as the CDF of $|e_D(k)|$ in fault-free condition. In this study, we selected the value $P_F = 10\%$.

## V. Directional Residual Design

In this effort, we have not used (for the fault isolation) the primary directional residuals that naturally originates in (9), instead, more effective residuals have been derived through the solution of an ad-hoc multi-objective optimization problem aimed at identifying the best fault signatures (column vectors of $W'$) that are mutually orthogonal and minimize the residual error evaluated on the experimental training data. These requirements are formalized in the following optimization problem:

$$\begin{aligned} &\min_{W_{Is} \in \mathbb{R}^{n_x \times n}} l \\ s.t. \ &\begin{cases} diag(W_{Is}) = -1 \\ \overline{w_{Is,i}}^T \overline{w_{Is,j}} = 0, \quad \forall i,j = 1,\ldots,n_x; \ i \neq j \\ \|Z W_{Is}^T\| < l \end{cases} \end{aligned} \quad (15)$$

where $W_{Is} \in \mathbb{R}^{n_x \times n}$ is the unknown optimized fault direction matrix to be computed, $diag(W_{Is})$ indicates the elements on the diagonal of $W_{Is}$, $Z$ is the matrix of the $m \times n$ experimental data, $w_{Is,i}$ is the $i$-th column vector of $W_{Is}$, and $\overline{w_{Is,i}}$ is its versor. The following considerations are in order:

- The constraint $diag(W_{Is}) = -1$ is structural, therefore the elements on the diagonal of $W_{Is}$ are set equal to $-1$;
- The constraint $\overline{w_{Is,i}}^T \overline{w_{Is,j}} = 0$ imposes the orthogonality between vectors $w_{Is,i}$ and $w_{Is,j}$;
- $Z W_{Is}^T \in \mathbb{R}^{m \times n_x}$ is the fault-free residual signal generated by the experimental data and $\|Z W_{Is}^t\|$ is the corresponding norm-2.

The resulting optimized residual vector generated by the fault signatures $W_{Is}$ is therefore

$$r_{Is}(k) = W_{Is} z(k) \quad (16)$$

By taking advantage of the directional property of $W_{Is}$, error information is derived by comparing the direction of the current residual vector $r_{Is}(k)$ with the known $n_x$ fault directions $w_{Is,i}$. In practice, the FI logic was implemented as described in the following section. As the first step, to be independent of the fault amplitude, both the residual $r_{Is}(k)$ and the sensor fault directions are normalized to unity norm. Then, the angular distance $d_i(k) = \angle(\overline{r_{Is}}(k), \overline{w_{Is,i}})$ between the normalized residual $\overline{r_{Is}}(k)$ and the normalized fault directions $\overline{w_{Is,i}}$ is computed.

## VI. The Recursive Evidence Based-Filter for Fault Isolation

In this section, it is introduced an evidence-based filter that is used to recursively combine the directional residue information providing a robust sensor fault isolation.

### A. Basic Belief Assignment (BBA)

Starting from the distances $d_i(k)$, $i = 1,\ldots,n_x$ and the error $e_D(k)$, the Basic Belief Masses (BBMs or masses) $m_k$ of the Frame of Discernment (FoD) $\Theta = \{F_1, \ldots, F_n, NF\}$ [12] ($F_i$ is the event "fault associated to the $i$-th sensor" and $NF$ is the "No-Fault" event) are assigned with the following rules.

If $|e_D(k)| > Th_D$, then:

$$m_k(i) = \begin{cases} \eta(2 - e^{\gamma d_i(k)}), & \forall i = 1,\ldots,n_x \\ \eta\left(1 - [1 + e^{\lambda(|e_D(k)| - Th_D)}]^{-1}\right), & i = n_x + 1 \end{cases} \quad (17)$$

If $|e_D(k)| \leq Th_D$, then:

$$m_k(i) = \begin{cases} [n_x(1 + e^{\lambda(|e_D(k)| - Th_D)})]^{-1}, & \forall i = 1,\ldots,n_x \\ 1 - [1 + e^{\lambda(|e_D(k)| - Th_D)}]^{-1}, & i = n_x + 1 \end{cases} \quad (18)$$

where $\gamma$ and $\lambda$ are free-design coefficients that control the slope of the BBA functions and $\eta$ is a scaling factor such that $\sum_{i=1}^{n_x+1} m_k(i) = 1$.

## B. Dempster–Shafer Combination Rule

A recursive Dempster–Shafer Combination Rule for managing the stream of information that originates from sensors measurements is proposed. The original (conjunctive) Dempster–Shafer Combination Rule of two evidences $m_1$ and $m_2$ is defined as [12]:

$$m_{1 \oplus 2}(A) = H^{-1} \sum_{B \cap C = A \neq \emptyset} m_1(B) \, m_2(C)$$
$$H = 1 - \sum_{B \cap C = \emptyset} m_1(B) \, m_2(C) \qquad (19)$$

It is well known [24] that combination rule (19) may generate, in some cases, paradoxical results that are emphasized in the case of sequential application of (19) over time. In particular (19) may cause the fast saturation of the belief function toward a "winning event" at the expense of the zeroing of the beliefs of all the remaining events with no possibility of self-desaturation [11]. Furthermore, the combination rule (19) does not allow to weight the current reliability level of the computed masses used for the combination.

## C. Reliability-based Dempster–Shafer Combination Rule

In the proposed recursive Dempster-Shafer Combination Rule, the reliability measure of the assigned masses is considered introducing a "Reliability signal" $Rel(k) \in [0,1]$, and a new recursive combination mechanism described in the following:

1) Given the posterior combined mass of an event previously calculated $m_{k-2 \oplus k-1}$ and the current measured evidence $m_k$ at sample time $k$, the original Dempster–Shafer combination of $m_{k-2 \oplus k-1}$ and $m_k$ is computed by applying (20):

$$m^{\#}_{k-1 \oplus k}(A) = H^{-1} \sum_{B \cap C = A \neq \emptyset} m_{k-2 \oplus k-1}(B) \, m_k(C)$$
$$H = 1 - \sum_{B \cap C = \emptyset} m_{k-2 \oplus k-1}(B) \, m_k(C) \qquad (20)$$

where $m^{\#}_{k-1 \oplus k}$ is the current prior combined mass.

2) The difference between $m^{\#}_{k-1 \oplus k}$ and $m_{k-2 \oplus k-1}$ is calculated and memorized in the vector $\Delta(k) \in \mathbb{R}^{|\Theta|=n_x+1}$:

$$\Delta(k) = m^{\#}_{k-1 \oplus k} - m_{k-2 \oplus k-1} \qquad (21)$$

3) The combined mass $m'_{k-1 \oplus k}$ at the sample time $k$ is computed as follows [25]:

$$m'_{k-1 \oplus k} = m_{k-2 \oplus k-1} + \Delta(k) \cdot Rel(k) \qquad (22)$$

Analyzing (22) it is evident that the posterior combined mass at time $k$, in case $Rel(k) = 0$, is equal to the posterior combined mass at the previous sample; in case $Rel(k) = 1$ this is updated and is equal to the prior combined mass at the current sample time $k$. In the intermediate case, the innovation is function of the reliability level $Rel(k)$. In this paper, the signal $Rel(k)$ is defined to be a function of the norm-2 of the control activity vector $u(k)$ using the following law:

$$Rel(k) = 1 - \left[1 + e^{\delta(Th_R - \|u(k)\|)}\right]^{-1} \qquad (23)$$

Where $\delta$ is a design coefficient that controls the slope of the sigmoidal function in (23) and $Th_R$ is a threshold that is derived from the experimental data as follows:

$$Th_R = min(\|u(k)\|) : CDF(\|u(k)\|) \geq (1 - P_F) \qquad (24)$$

where $P_F = 10\%$.

The motivation of using $\|u(k)\|$ as the reliability signal is based on the consideration that in case of aircraft maneuvers (significant control activity) the modeling uncertainties increases and therefore the information provided by the residual signal is less reliable (the reliability signal is low). In addition, it is also introduced the following practical de-saturation mechanism to avoid the locking of the mass $m'_{k-1 \oplus k}$ at the boundary values ($mass = 0$ or $mass = 1$).

$$m''_{k-1 \oplus k}(A) = min[m'_{k-1 \oplus k}(A), 10^{-4}] \qquad (25)$$

$$m_{k-1 \oplus k} = \chi^{-1} m''_{k-1 \oplus k}$$
$$\chi = \sum_{A \in \Theta} m''_{k-1 \oplus k}(A) \qquad (26)$$

## D. Fault Isolation logic

Under the assumption of normal operation or in case of a single sensor fault, a simple logic can be employed for FI, that is: the event in the Frame of Discernment with the largest posterior mass $m_{k-1 \oplus k}$ is the most probable event. Specifically, for each $k$, if the maximum posterior mass is associated to a sensor, then this sensor is considered the fault one, otherwise, if the largest $m_{k-1 \oplus k}$ is assigned to the No-Fault, then we are in normal fault-free condition.

## VII. AIRCRAFT AND FLIGHT DATA

The proposed technique has been designed and validated using flight data of a Tecnam P92 aircraft [26]. The aircraft mass is approximately 600 Kg and the propulsion is provided by a 74 kW Rotax 912 ULS with a two-blade fixed-pitch propeller, for a maximum cruise speed of 219 km/h and an operational ceiling of 4200 m. Data were acquired in semi-autonomous mode, that is, the aircraft was manually flown by a pilot during the take-off, and landing phase and flown autonomously in cruise flight condition. A set of six-flight normalized datasets (null mean value and unit variance) was considered in this study. Five flights (1 hour and 20 minutes) were used for the design and the remaining flight (16 minutes) for validation of the overall scheme. Data sampling time is 0.1 s. The study does not consider data associated with take-off, initial climb, final descent, and landing phases. The set of 12 signals listed in Table I were considered in this study [27], of which the first 8 are the monitored sensors $x(k)$ while the last 4 are the actuation signals $u(k)$. All the signals have been normalized to zero mean ad unitary variance.

TABLE I. AIRCRAFT SENSORS

| $x(k)$ | | | | $u(k)$ | |
|---|---|---|---|---|---|
| α | Angle of attack | P | Roll speed | Alt | Altitude |
| β | Drifting angle | Q | Pitch speed | Aie | Aileron |
| TaS | True AirSpeed | R | Yaw speed | Rud | Rudder |
| φ | Roll angle | θ | Pitch angle | Thr | Thrust lever |

## VIII. OFFLINE DESIGN OF THE FAULT DIAGNOSIS SYSTEM

In this section, it is described the design of the different blocks that constitute the scheme in Figure 1.

### A. Fault Detection block design

The FD vector $v$ to be used in (12) was derived exploiting the aircraft training data set, resulting in:

$v^T = 10^{-2}[69 \quad -1 \quad 67 \quad -9 \quad -11 \quad 10 \quad -8 \quad 3 \quad 9 \quad 6 \quad -17 \quad 4]$

while the experimental detection threshold computed using (14) resulted equal to $Th_D = 0.24$.

### B. Fault Isolation block design

The value of coefficients that define the BBA in (17) is set to $\gamma = ln(2)/90$ and $\lambda = -20\,ln(3)/Th_D$. These values were selected such that in the case the angular distance is maximum (90 deg), then $m_k(i)$ is null; if $|e_D(k)| = 0.9\,Th_D$, then $m_k(n_x + 1) = 0.9\eta$ and if $|e_D(k)| = 1.1\,Th_D$, then $m_k(n_x + 1) = 0.1\eta$. The Reliability threshold was computed applying (24) resulting in $Th_R = 2.43$ while the slope coefficient $\delta = 40\,ln(3)/Th_R$ was fixed for the signal $Rel(k)$ so that, if $\|u(k)\| = 0.95\,Th_R$, then $Rel(k) = 0.9$; and if $\|u(k)\| = 1.05\,Th_R$, then $Rel(k) = 0.1$.

## IX. RESULTS

The analysis was performed considering a 16-minute validation flight and injecting single sensor faults having a rectangular shape of 12-minute duration. For each one of the 8 monitored sensors, quantitative Fault Detection and Isolation performance for the proposed diagnostic system are compared to other State-of-the-Art (SoA) schemes. Due to space limitations, it was not possible to report the results for all the techniques and studies. The following acronyms were used to identify the different Fault isolation schemes. "RB" refers to our proposed FDI system based on the Reliability-Based Dempster-Shafter combination rule, and "DS", "PCR6", and "WE" refer to state-of-the-art techniques based on the canonical Dempster–Shafer combination rule [12], Proportional Conflict Redistribution rule no. 6 [13], and Weighted Evidence Combination Rule [11], respectively. The fault amplitudes used for the quantitative validation were derived from the training data and were set equal to 3 times the (rounded) mean estimation error obtained in case of the basic LS model in (8) and are listed in the first row of Table II. The first considered performance metric was the True Detection Rate (TDR) that measures the % ratio between the number of faulty samples correctly detected by the FD block and the number of samples the fault is actually active (see Table II). Table III reports the True Isolation Rate (TIR) that measures the % ratio between the number of samples the FI block correctly isolates the faulty sensor and the number of samples that the sensor is actually faulty.

TABLE II. TRUE DETECTION RATE [%]

| Fault amplitude | α [4°] | β [4°] | TaS [3m/s] | P [3°/s] | Q [6°/s] | R [6°/s] | φ [9°] | θ [6°] |
|---|---|---|---|---|---|---|---|---|
| RB | **96** | **100** | **94** | **97** | 72 | **96** | 86 | **98** |
| WE | 89 | **100** | 86 | 91 | 74 | 89 | 91 | 93 |
| PCR6 | 88 | 99 | 85 | 90 | 76 | 88 | 90 | 93 |
| DS | 90 | **100** | 87 | 92 | **77** | 90 | **92** | 94 |

TABLE III. TRUE ISOLATION RATE [%]

| Fault amplitude | α [4°] | β [4°] | TaS [3m/s] | P [3°/s] | Q [6°/s] | R [6°/s] | φ [9°] | θ [6°] |
|---|---|---|---|---|---|---|---|---|
| RB | **88** | **98** | **93** | **91** | 72 | **96** | **74** | **91** |
| WE | 79 | 1 | 74 | 67 | 68 | 77 | 66 | 70 |
| PCR6 | 71 | 88 | 77 | 70 | **76** | 82 | 67 | 79 |
| DS | 76 | 89 | 78 | 73 | **76** | 83 | 68 | 80 |

Analyzing Tables II and III it is apparent that the proposed approach performs better for most sensors especially at FI level. To deepen the analysis, time-domain results are shown to illustrate the operation of the FDI system in case of a rectangular sensor fault on the sensor $TaS$.

In Figure 2 the FD signal $|e_D(k)|$ and the FD threshold $Th_D$ are shown. The red line indicates the fault shape (rectangular fault in the interval between 2 and 14 minutes). It is observed that the FD signal is, most of the time, under the threshold in the fault-free time intervals, while it crosses the FD threshold when the fault is active. It is also observed the presence of sporadic false detections and missed detections.

Following the computation of the angular distance between the residual $r(k)$ and the $n_x$ optimized fault directions, the BBMs in (17) and (18) are computed for all the event in the FoD, while $Rel(k)$ in (23) is calculated as a function of the actuation signals (see Figure 3). It is observed that the mass of the $NF$ event is large when there is no failure, and small when the fault is present. Since the sum of all the masses must always be equal to 100%, the instantaneous BBM for the $TaS(k)$ sensor does not dominate over the others (when the fault is active) making fault isolation based on BBMs a complex task. On the other side, combining the evidences and the reliability signal with the proposed DS Combination Rule produces the combined masses shown in Figure 4, where it is now evident that the faulty sensor is $TaS(k)$.

During the first minute of flight and after the fault injection (14-16 minutes), some samples exceed the Detection threshold $Th_D$, see Figure 2. These were considered false alarms (and correctly not detected, see Figure 3) because the reliability signal was low, and the evidence was consistently accumulated. Similarly, the samples between 2 and 14 minutes that are under the $Th_D$ do not produce false negatives thanks to the evidence accumulation see Figure 4.

The utility of the reliability signal in the recursive combination rule is clarified in Figure 5, where the evolution of the fault mass associated to the $TaS(k)$ sensor for the different combination methods is compared (30s period during the 13-th minute). It is observed that, while the other SoA combination rules are influenced by noise and aircraft maneuvers, the proposed combination rule, during a maneuver, blocks the evolution of the mass preventing the rapid decrease of the fault evidence associated to the TaS sensor. The result is that the scheme correctly isolates the fault on TaS, while the other techniques fail the correct isolation. Around 13:25 the reliability signal decreases and the proposed combination technique decreases the evidence on the TaS sensor fault, nevertheless as shown in Figure 4 the evidence on the TaS sensor keeps higher than the accumulated evidence of other events and therefore the Isolation is again correct most of the time.

## X. CONCLUSION

A robust data-driven Fault Diagnosis scheme based on a recursive DS combination rule has been presented. The study shows that the proposed reliability-based DS combination rule provides superior performance compared to state-of-art recursive DS combination rules. Experiments based on real flight data highlight that it is possible to achieve 93% correct FI for a fault amplitude of 3 m/s on the True Air Speed sensor. In general, results shows a substantial increment of the TIR index for all the 8 monitored sensors of 10-12% compared to SoA methods.

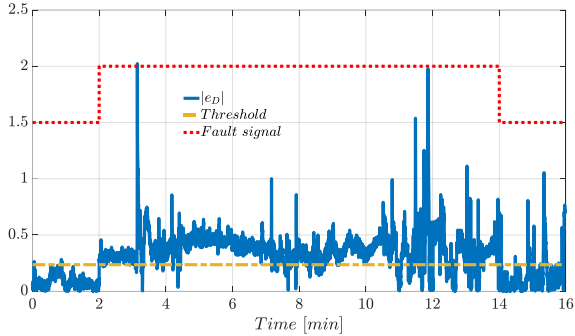

Figure 2. *Fault Detection signals. Fault on the $TaS(k)$ sensor.*

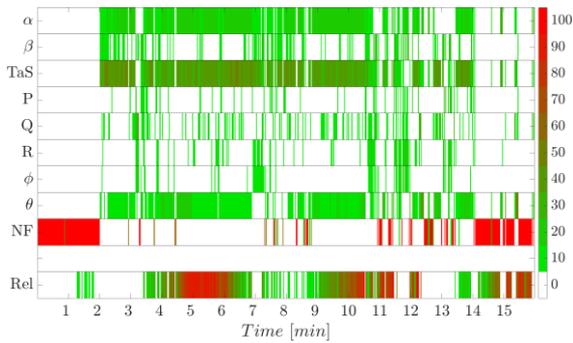

Figure 3. *BBMs and the reliability signal $Rel(k)$ as a function of the time (the fault on $TaS(k)$ is active in the interval time [2-14] m). The color scale defines the value of the mass and reliability signal.*

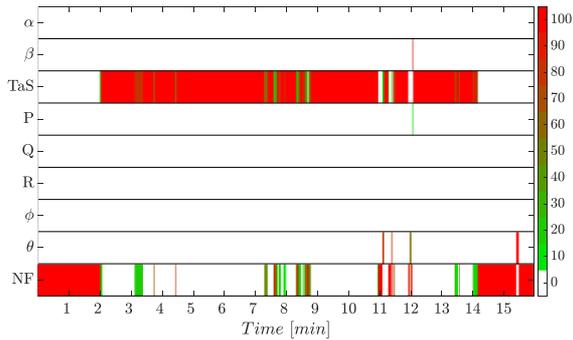

Figure 4. *Combined masses as function of time. Fault on the $TaS(k)$ sensor*

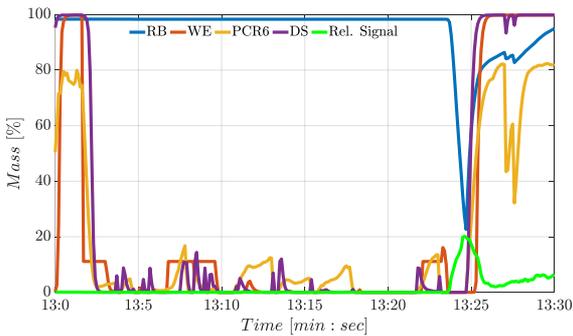

Figure 5. *Evolution of fault evidence on the TaS sensor for the different combination rules.*